\documentclass[twocolumn,english,prl,a4paper]{revtex4}
\usepackage[T1]{fontenc}
\usepackage[latin9]{inputenc}
\usepackage{amsmath}
\usepackage{graphicx}
\usepackage{amssymb}
\usepackage{esint}
\usepackage{color}
\usepackage[letterpaper]{geometry}
\makeatletter
\@ifundefined{textcolor}{}
{%
 \definecolor{BLACK}{gray}{0}
 \definecolor{WHITE}{gray}{1}
 \definecolor{RED}{rgb}{1,0,0}
 \definecolor{GREEN}{rgb}{0,1,0}
 \definecolor{BLUE}{rgb}{0,0,1}
 \definecolor{CYAN}{cmyk}{1,0,0,0}
 \definecolor{MAGENTA}{cmyk}{0,1,0,0}
 \definecolor{YELLOW}{cmyk}{0,0,1,0}
 }

\makeatother

\usepackage{babel}

\begin{document}

\title{Universal Markovian reduction of Brownian particle dynamics}

%\author{R. Martinazzo}
%\email{rocco.martinazzo@unimi.it}
%\affiliation{Dipartimento di Chimica-Fisica ed Elettrochimica, Università degli
%Studi di Milano, v. Golgi 19, 20133 Milano, Italy}
%\author{B. Vacchini}
%\affiliation{\mbox{ $^{a}$Dipartimento di Fisica, Università degli Studi di
%Milano, v. Celoria 16, 20133 Milano, Italy}\\
% \mbox{$^{b}$INFN, Sezione di Milano, v. Celoria 16, 20133 Milano,
%Italy} }
%\author{K. Hughes}
%\affiliation{School of Chemistry, University of Wales Bangor, Bangor, Gwynedd
%LL57 2UW, United Kingdom }
%\author{I. Burghardt}
%\affiliation{Département de Chimie, Ecole Normale Supérieure, 24 rue Lhomond,
%F-75231 Paris, France }

\author{R. Martinazzo$^{a}$\footnote{rocco.martinazzo@unimi.it}, B. Vacchini$^{b,c}$, K. Hughes$^{d}$ and I. Burghardt$^{e}$}
\affiliation{$^{a}$Dipartimento di Chimica-Fisica ed Elettrochimica, Università degli
Studi di Milano, v. Golgi 19, 20133 Milano, Italy}
\affiliation{$^{b}$Dipartimento di Fisica, Università degli Studi di
Milano, v. Celoria 16, 20133 Milano, Italy}
\affiliation{$^{c}$INFN, Sezione di Milano, v. Celoria 16, 20133 Milano,
Italy }
\affiliation{$^{d}$School of Chemistry, University of Wales Bangor, Bangor, Gwynedd
LL57 2UW, United Kingdom }
\affiliation{$^{e}$Département de Chimie, Ecole Normale Supérieure, 24 rue Lhomond,
F-75231 Paris, France }

\begin{abstract}
Non-Markovian processes can often be turned Markovian by enlarging the set of
variables. Here we show, by an explicit construction, how this can be done for
the dynamics of a Brownian particle obeying the generalized Langevin 
equation. Given an arbitrary bath spectral density $J_{0}$, we
introduce an orthogonal transformation of the bath variables into effective
modes, leading stepwise to a semi-infinite chain with nearest-neighbor
interactions. The transformation is uniquely determined by $J_{0}$ and
defines a sequence $\{J_{n}\}_{n\in\mathbb{N}}$ of residual spectral
densities describing the interaction of the terminal chain mode, at each step, 
with the remaining bath. We derive a simple, one-term recurrence relation for this sequence, 
and show that its limit is the quasi-Ohmic expression provided by the Rubin model of
dissipation. Numerical calculations show that, irrespective of the details of
$J_{0}$, convergence is fast enough to be useful in practice for an effective
Markovian reduction of quantum dissipative dynamics.
\end{abstract}
\maketitle
\emph{Introduction.} As is well known, the study of open systems, in
both the classical and quantum case, is a subject of major 
interest in physics, chemistry, and various other disciplines. 
In many applications
and fundamental experiments, one is faced with 
the reduced dynamics of a relatively simple subsystem which can be
manipulated and measured, while the environment is only partially
under control. A thorough understanding of the ensuing dynamics has
been obtained for the Markovian case, in which feedback from the environment
to the system can be neglected, and general analytical results
are available together with efficient numerical algorithms \citep{Alicki2007,Breuer2007,Weiss2008}.
The situation is much more involved
in the non-Markovian regime, which typically arises 
due to strong coupling and similar time scales of 
system and bath evolution. 
In this case general strategies are still available
\citep{Breuer2007,Weiss2008}, but differently from the Markovian
case they typically lack simple results of general validity, to be
expressed in terms of the phenomenologically relevant quantities and
leading to manageable numerical tasks. A bridge between the 
two situations
can be built relying on a suitable embedding of a non-Markovian dynamics
in a Markovian one, as recently addressed in \citep{Hughes2009a,Hughes2009b,Siegle2010a}.
Indeed, while it is common wisdom that a non-Markovian process can
be embedded in a Markovian one by a suitable enlargement of the number
of relevant variables already at classical level \citep{Cox1965},
there is no universal recipe for how this can be done and which class
of non-Markovian processes can be reached.

In the present Letter we demonstrate how such a Markovian reduction can
be achieved for the ubiquitous model of quantum dissipation provided
by a Brownian particle, or a two-level system, linearly coupled to
a bath of harmonic oscillators characterized by an arbitrary spectral
density (SD) \citep{Caldeira1983a}. The procedure is physically transparent,
in that it focuses exclusively on the SD, and all relevant quantities 
can be constructed in terms of the SD. As will be shown below, 
the system dynamics can equivalently be described including, besides the Brownian
particle degree of freedom, a set of effective environmental modes
coupled in a linear-chain fashion. The terminal mode of the chain couples to a residual bath and undergoes a Brownian-like dynamics which rapidly approaches a Markovian behavior over the whole interval of relevant frequencies as the length of the chain increases. 
The model as such is closely related to Mori's theory \citep{mori} and its generalizations \citep{dupuis,grigolini}.  
While previous work by two of us \citep{Hughes2009a,Hughes2009b} 
has focused on the implications of a Markovian truncation of such effective mode
chains, the present analysis proves the convergence towards Ohmic 
behavior, and thus the general validity of the procedure. 
The question of how to correctly set the initial state of the chain
will be detailed in a forthcoming paper. 

\emph{Effective-mode transformation.} We start by considering the
Caldeira-Leggett Hamiltonian, here written in mass-weighted bath coordinates
$x_{k}$,
\begin{equation}
H=\frac{p^{2}}{2m}+V(s)+\frac{1}{2}\sum_{k=1}^{N}\left[p_{k}^{2}+\omega_{k}^{2}\left(x_{k}-\frac{c_{k}}{\omega_{k}^{2}}s\right)^{2}\right]\label{eq:GLE Hamiltonian}\end{equation}
which is known to lead, in the continuum limit, to a generalized Langevin
dynamics for the system described by the $s$ degree of freedom. The
reduced system dynamics is entirely determined by the SD of the environmental
coupling $J_{0}(\omega)$ which, for the microscopic model above,
reads as \citep{Weiss2008} \begin{equation}
J_{0}(\omega)=\frac{\pi}{2}\sum_{k=1}^{N}\frac{c_{k}^{2}}{\omega_{k}}\delta(\omega-\omega_{k}).\label{eq:discrete spectral density}\end{equation}
In general, $J_{0}(\omega)$ is a real, odd parity function defined by the
real part of the frequency-dependent memory kernel %
\footnote{Here and in the following we define the Fourier transform as $\gamma(\omega)=\int_{-\infty}^{+\infty}\gamma(t)e^{i\omega t}dt$.%
} $\gamma(\omega)$ entering the generalized Langevin equation (GLE),
namely $J_{0}(\omega)=m\omega\mbox{Re}\gamma(\omega)$ and $J_{0}(\omega)\geqslant0$
for $\omega>0$. It fully determines $\gamma(\omega)$ by virtue of
the Kramers-Kronig relations, as well as the correlation function
of the GLE random force by virtue of the fluctuation-dissipation theorem.
In the following we assume, as a typical situation, that $J_{0}(\omega)$
is strictly positive and continuous in an interval $(0,\omega_{R})$
- where $\omega_{R}$ is a high-frequency cutoff - and zero otherwise
on the positive real axis; other interesting cases will be briefly
considered below. 

\begin{figure}
\begin{centering}
\includegraphics[width=1\columnwidth]{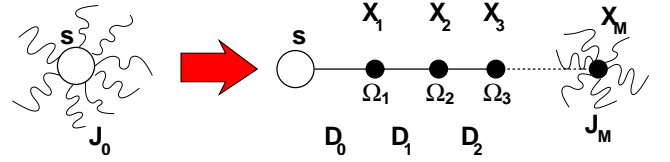}
\par\end{centering}
\caption{\label{fig:Schematics}(Color online) Schematic for the linear-chain transformation
described in the text. $X_{n}$ are the effective modes, $\Omega_{n}$
their frequencies and $D_{n}$ the couplings between adjacent modes
$X_{n-1}$ and $X_{n}$ ($X_{0}=s$). The SD $J_{M}$ ($M\geq0$)
describe the interaction with the residual bath. }
\end{figure}

Given a GLE and its relevant SD $J_{0}$, Eq.(\ref{eq:discrete spectral density})
allows to define a microscopic model for the dissipative dynamics
of the $s$ degree of freedom, \emph{e.g.} by introducing a bath of
harmonic oscillators with evenly spaced frequencies $\omega_{k}=k\Delta\omega$
($k=1,\ldots,N$) and setting the coupling coefficients of Eq.(\ref{eq:GLE Hamiltonian})
as \begin{equation}
c_{k}=\sqrt{\frac{2\omega_{k}\Delta\omega J_{0}(\omega_{k})}{\pi}} \label{eq:coeffs}\end{equation}
The system-bath interaction term
in Eq.(\ref{eq:GLE Hamiltonian}), $H^{int}=-\sum_{k=1}^{N}c_{k}x_{k}s=-D_{0}X_{1}s$,
naturally introduces an effective mode $X_{1}=\sum_{k=1}^{N}c_{k}x_{k}/D_{0}$
where $D_{0}$ is a normalization constant which in the continuum
limit reads $D_{0}^{2}=\sum_{k=1}^{N}c_{k}^{2}\approx\frac{2}{\pi}\int_{0}^{\infty}d\omega J_{0}(\omega)\omega$.
This defines the first column of an otherwise arbitrary, orthogonal
matrix $\boldsymbol{T}$ transforming the bath coordinates $\boldsymbol{x}^{t}=(x_{1},\ldots,x_{N})$
into $\boldsymbol{X}^{t}=(X_{1},X'_{2},\ldots,X'_{N})$, $\boldsymbol{X}=\boldsymbol{T}^{t}\boldsymbol{x}$.
The transformation can be fixed by requiring that the {}``residual''
bath of coordinates $X'_{2},X'_{3},\ldots,X'_{N}$ is in normal form,
\emph{i.e.} that $(\boldsymbol{T}^{t}\boldsymbol{\omega^{2}}\boldsymbol{T})_{ij}=\delta_{ij}\bar{\Omega}_{i}^{2}$
holds for all $i,j>1$ with $(\boldsymbol{\omega^{2}})_{ij}=\delta_{ij}\omega_{i}^{2}$
($i,j=1,\ldots,N$) . Note that the frequency $\Omega_{1}$ of the
effective mode introduced in this way is solely determined by the
SD $J_{0}(\omega)$, $\Omega_{1}^{2}=(\boldsymbol{T}^{t}\boldsymbol{\omega^{2}}\boldsymbol{T})_{11}=\sum_{k=1}^{N}\omega_{k}^{2}c_{k}^{2}/D_{0}^{2}\thickapprox\frac{2}{\pi D_{0}^{2}}\int_{0}^{\infty}d\omega J_{0}(\omega)\omega^{3}$.
The couplings $C_{k}=-(\boldsymbol{T}^{t}\boldsymbol{\omega^{2}}\boldsymbol{T})_{1,k}$
($k=2,\ldots,N$) between the normal modes 
of the residual bath and $X_{1}$ allow one to re-write Eq.(\ref{eq:GLE Hamiltonian})
as a Caldeira-Leggett-like Hamiltonian for the $X_{1}$ degree of
freedom, thereby defining the SD 
$J_{1}(\omega)=\frac{\pi}{2}\sum_{k=2}^{N}\frac{C_{k}^{2}}{\bar{\Omega}_{k}}\delta(\omega-\bar{\Omega}_{k})$
{}``felt'' by the effective mode $X_{1}$, which is the only bath
mode directly coupled to the $s$ degree of freedom. Clearly, in the
continuum limit, the procedure can be indefinitely iterated and used
to define a sequence of effective modes $X_{1},X_{2},\ldots,X_{M},\ldots$
coupled in a linear-chain fashion \emph{and} a corresponding sequence
of SD $J_{1},J_{2},\ldots,J_{M},\ldots$ characterizing the residual
bath {}``felt'' by each mode, see Fig.\ref{fig:Schematics}. 
In other words, there exists an orthogonal coordinate transformation
which converts the continuum version of Eq.(\ref{eq:GLE Hamiltonian})
into the form\begin{equation}
\begin{array}{cc}
H & =\frac{p^{2}}{2m}+V(s)+\Delta V(s)-D_{0}sX_{1}+\\
\\ & -\sum_{n=1}^{\infty}D_{n}X_{n}X_{n+1}+\frac{1}{2}\sum_{n=1}^{\infty}\left[P_{n}^{2}+\Omega_{n}^{2}X_{n}^{2}\right]\end{array}\label{eq:EMH}\end{equation}
where, for $n\geqslant0$,
\[D_{n}^{2}=\frac{2}{\pi}\int_{0}^{\infty}d\omega J_{n}(\omega)\omega,\,\,\, 
\Omega_{n+1}^{2}=\frac{2}{\pi D_{n}^{2}}\int_{0}^{\infty}d\omega J_{n}(\omega)\omega^{3}\]
%\begin{equation}
%\Omega_{n+1}^{2}=\frac{2}{\pi D_{n}^{2}}\int_{0}^{\infty}d\omega J_{n}(\omega)\omega^{3},\label{eq:omegan}\end{equation}
\[X_{n+1}=\sqrt{\frac{2}{\pi D_{n}^{2}}}\int_{0}^{\infty}\omega\sqrt{J_{n}(\omega)}x(\omega)d\omega\]
(and similarly for $P_{n+1}$ in terms of $p(\omega)$); $\Delta V(s)=\delta\Omega_{0}^{2}s^{2}/2$
is a counter term involving the renormalization frequency $\delta\Omega_{0}^{2}=(2/\pi)\int_{0}^{\infty}d\omega J_{0}(\omega)/\omega$.
%Notice that although different canonical transformations to a linear
%chain can be devised, \emph{e.g.} starting from mass-and-frequency
%weighted bath coordinates \citep{cederbaum05}, the one suggested above is 
%entirely determined
%by the sequence $\{J_{n}\}_{n\in\mathbb{N}}$. Conversely, as we show
%below, since the coupling in Eq.(\ref{eq:EMH}) is of pure coordinate
%form, we can write down an explicit expression for such sequence \emph{without}
%knowing the eigenfrequencies of the residual bath at each step. 
As we show below, though different canonical transformations to a linear chain be devised (see e.g. Ref. \onlinecite{cederbaum05}), the one suggested above, with the coupling in pure coordinate form, allows one to write an explicit expression for $\{J_{n}\}_{n\in\mathbb{N}}$ \emph{without} knowing the eigenfrequencies of the residual bath at each step.

\emph{Sequence of SD.} As observed by Leggett \citep{Leggett1984a,Garg1985a},
the SD acting on the system degree of freedom can be obtained from the
analytically continued, Fourier-transformed classical (or Heisenberg)
equations of motion as the imaginary part of a propagator,
\[
J_{0}(\omega)=-\lim_{\epsilon\rightarrow0^{+}}\mbox{Im}L_{0}(\omega+i\epsilon)\equiv-\mbox{Im}L_{0}^{+}(\omega).\]
This procedure has recently been used by two of us
to obtain a continued-fraction
expression for the SD generated by a linear chain with Markovian closure,
which in turn formed the basis for approximating a given SD 
\citep{Hughes2009a,Hughes2009b}. Employing a similar strategy, we now focus on the properties of the 
residual spectral densities $J_{M}$ closing the chain after
$M$ effective modes have been extracted as outlined above.
For the Hamiltonian of Eq.(\ref{eq:GLE Hamiltonian}), after introducing the
first effective mode $X_{1}$, we obtain \[
L_{0}(z)=-z^{2}-\cfrac{D_{0}^{2}}{\Omega_{1}^{2}-z^{2}-\sum_{k}\cfrac{C_{k}^{2}}{\bar{\Omega}_{k}^{2}-z^{2}}}\]
where $\bar{\Omega}_{k}$ and $C_{k}$ have been introduced above.
In the continuum limit, with the help of Eq.(\ref{eq:coeffs}), the sum in the denominator can be replaced by the function 
\footnote{This is legitimate since the (unknown) eigenfrequencies $\bar{\Omega}_{k}$
satisfy $\omega_{1}\leq\bar{\Omega}_{2}\leq\omega_{2}...\leq\bar{\Omega}_{N}\leq\omega_{N}$,
thereby covering uniformly the interval $(0,\omega_{R})$ as $\Delta\omega\rightarrow0$.%
} \begin{equation*}
W_{1}(z) = \sum_{k=2}^{N}\frac{C_{k}^{2}}{\bar{\Omega}_{k}^{2}-z^{2}}\approx
 \frac{2}{\pi}\int_{0}^{\infty}d\omega\frac{J_{1}(\omega)\omega}{\omega^{2}-z^{2}}\end{equation*}
or, equivalently,  
\begin{equation}
W_{1}(z)=\frac{1}{\pi}\int_{-\infty}^{+\infty}d\omega\frac{J_{1}(\omega)}{\omega-z}.\label{eq:Cauchy}\end{equation}
In this form $W_{1}$ is given as an integral of its
limiting imaginary part, $J_{1}(\omega)=\mbox{Im}W_{1}^{+}(\omega)$.
In the following, a function $W_{1}$ defined by Eq.(\ref{eq:Cauchy})
will be referred to as the \emph{Cauchy transform} of $J_{1}$ \citep{Muskhelishvili1953};
it is an analytic function in the whole complex plane except for the
support of $J_{1}$ on the real axis %
\footnote{In the presence of a cutoff $\omega_{R}$ the upper and lower half
planes are connected through the semiaxes $|\omega|>\omega_{R}$.
Note also that we consider only Cauchy transforms of odd functions. %
}, which vanishes as $z^{-2}$ for $|z|\rightarrow\infty$. We \emph{define}
\begin{equation}
W_{0}(z)=\frac{D_{0}^{2}}{\Omega_{1}^{2}-z^{2}-W_{1}(z)}\label{eq:W0-def}\end{equation}
which, analogously to $L_{0}=-z^{2}-W_{0}$, gives $J_{0}(\omega)=\mbox{Im}W_{0}^{+}(\omega)$.
It follows that $J_{1}$ can be written in terms of $J_{0}$ as 
$J_{1}(\omega)=\frac{D_{0}^{2}J_{0}(\omega)}{|W_{0}^{+}(\omega)|^{2}}$
where, as we now show, $W_{0}$ is the Cauchy transform of $J_{0}$.
In order to prove this, we notice that according to its definition,
Eq.(\ref{eq:W0-def}), $W_{0}$ is analytic in the upper and lower
half planes %
\footnote{The denominator vanishes on the real axis only, since $\Omega_{1}^{2}-z^{2}-W_{1}(z)=0$
is the eigenvalue equation defining the frequencies $\omega_{k}$
appearing in Eq.(\ref{eq:GLE Hamiltonian}). %
}, and vanishes as $z^{-2}$ for $|z|\rightarrow\infty$. Writing $W_{0}(z)$
as a Cauchy integral on a large semicircle in the upper half plane,
we can add a term $\pm(\omega-z^{*})^{-1}$ to the integrand, and
get, from the real and imaginary parts of the resulting expression,
the desired result. In general, then \begin{equation}
J_{n+1}(\omega)=\frac{D_{n}^{2}J_{n}(\omega)}{|W_{n}^{+}(\omega)|^{2}}\label{eq:recursive-J*}\end{equation}
defines a recurrence relation for the SD $J_{n+1}(\omega)$ felt by
the $n+1$-th effective mode, given the SD $J_{n}(\omega)$ of the
$n$-th mode. Equivalently\footnote{According to Eq.(\ref{eq:recursive-J*}) the Cauchy transform
of $J_{n+1}$ must be of the form $W_{n+1}(z)=g_{n+1}(z)-\frac{D_{n}^{2}}{W_{n}(z)}$
where $g_{n+1}(z)$ is an analytic function with vanishing imaginary
part on the real axis, uniquely fixed by asking that it offsets the
behavior of $W_{n}^{-1}(z)$ as $|z|\rightarrow\infty$, $
W_{n}(z)\approx-\frac{D_{n}^{2}}{z^{2}}\left(1+\frac{\Omega_{n+1}^{2}}{z^{2}}+...\right).$
}, \begin{equation}
W_{n+1}(z)=\Omega_{n+1}^{2}-z^{2}-\frac{D_{n}^{2}}{W_{n}(z)}\label{eq:recursive-W*}\end{equation}
is a recurrence relation for the Cauchy transforms which only requires
the first Cauchy transform ($W_{0}$) as an input and easily provides 
the sequence $J_{n}(\omega)=\mbox{Im}W_{n}^{+}(\omega)$.

Eq.(\ref{eq:recursive-W*}) represents the main result of this Letter. It is  
a simple recurrence relation between the Cauchy transforms of the SDs 
which allows us to write the limiting condition as 
\[
W(z)=\Omega^{2}-z^{2}-\frac{D^{2}}{W(z)}\]
provided $\Omega=\lim_{n}\Omega_{n}$ and $D=\lim_{n}D_{n}$ exist.
The physical solution ($\mbox{Im}W^{+}\geqslant0$) 
provides the SD {}``closing'' the chain, 
which has a non-vanishing value for $\omega^{2}\in[\Omega^{2}-2D,\Omega^{2}+2D]$ only.
In other words, the limiting SD reads as
\begin{equation}
J(\omega)=\frac{1}{2}\sqrt{(\omega^{2}-\omega_{L}^{2})(\omega_{R}^{2}-\omega^{2})}\,\,\,\,\omega_{L}\leq\omega\leq\omega_{R}\label{eq:homogeneous chain}\end{equation}
where $\omega_{R}^{2}=\Omega^{2}+2D$ and $\omega_{L}^{2}=Max\{\Omega^{2}-2D,0\}$.
The requirement $J_{0}(\omega)>0$ for $\omega>0$ fixes $\Omega^{2}=2D$,
since the condition $\Omega^{2}>2D$ (though physically admissible)
would give rise (using the above recursion procedure backwards) to
a SD $J_{0}$ with a low-frequency cutoff $\omega_{L}=\sqrt{\Omega^{2}-2D}$.
Therefore 
\footnote{When $\Omega^{2}\leq2D$, $D$ and $\Omega$ are functions of $\omega_{R}$ only (see Eq.(\ref{eq:homogeneous chain})). 
It follows, from their definition, $\Omega^{2}=2D$.}, 
$\Omega^{2}=2D,\,\omega_{R}^{2}=4D=2\Omega^{2}$ and Eq.(\ref{eq:homogeneous chain})
reduces to the quasi-Ohmic SD provided by the Rubin model of dissipation
\citep{Rubin1963a,Weiss2008}, \begin{equation}
J_{Rubin}(\omega)=\frac{\omega\omega_{R}}{2}\sqrt{1-\frac{\omega^{2}}{\omega_{R}^{2}}}\Theta(\omega_{R}-\omega).\label{eq:Rubin}\end{equation}
This means that provided a sufficient number of effective modes is
included in the definition of the system, the resulting dynamics is
Markovian. In practice, as we show numerically below this number is
rather small, since convergence is quite fast
even for structured spectral densities. Notice though that when $J_{0}(\omega)$
has a low frequency cutoff $\omega_{L}$ but is otherwise positive
on the interval $(\omega_{L},\omega_{R})$, 
Eq.(\ref{eq:homogeneous chain}) shows that no Markovian reduction is possible, no matter
how many effective modes are included in the system. 
\begin{figure}
\noindent \begin{centering}
\includegraphics[width=0.9\columnwidth]{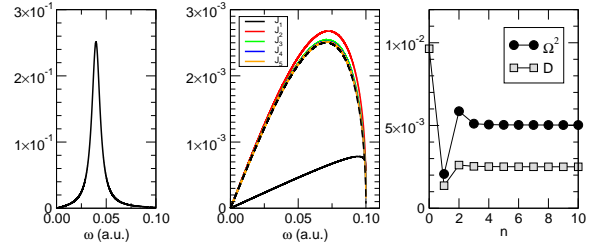}
\par\end{centering}
\caption{\label{fig:Garg}(Color online) Left: the SD $J_{0}$ defined in Eq.(\ref{eq:Garg})
for $\omega_{0}=0.04$ a.u., $d_{0}=0.01$ a.u. and $\gamma=0.01$
a.u.. Middle: results of the deconvolution of $J_{0}$ for the first
five modes, obtained when setting the high-frequency cutoff $\omega_{R}$
to $0.1$ a.u.. The Rubin SD of Eq.(\ref{eq:Rubin}) with the same
$\omega_{R}$ is shown as dashed line. Right: Effective modes parameters
($\Omega_{n}^{2}$ and $D_{n}$) up to $n=10$. }
\end{figure}

In general, a high-frequency cutoff $\omega_{R}$ can be naturally
associated to the SD $J_{0}(\omega)$, determining the spectrum of
environmental frequencies relevant for the reduced system's dynamics.
This suffices to show that $\mbox{Re}W_{0}^{+}(\omega)$ diverges logarithmically
for $\omega\rightarrow\omega_{R}$, unless $J_{0}(\omega_{R})$ is
equal to zero. In view of Eq.(\ref{eq:recursive-J*})
this in turn implies that $J_{1}(\omega)\rightarrow0$ as $\omega\rightarrow\omega_{R}$,
and this cutoff is later on automatically preserved. Furthermore,
starting from Eq.(\ref{eq:recursive-W*}) one can also immediately
obtain some interesting bounds on the value $\mbox{Re}W_{n}^{+}(\omega)$
can take at the extreme points of the relevant frequency interval $(0,\omega_{R})$,
which help determine the behavior of the recurrence relation Eq.(\ref{eq:recursive-J*}).
Among these we only note here that  
$\mbox{Re}W_{n}^{+}(\omega)$ is positive for $\omega=0$ and negative
for $\omega=\omega_{R}$, and then by continuity the function will go through
zero at some intermediate point $\bar{\omega}$. This explains why
the procedure generally fails to converge for SD with gaps, 
since if $\bar{\omega}$ falls in the gap one has $W_{n}^{+}(\bar{\omega})=0$
and, by virtue of Eq.(\ref{eq:recursive-W*}), this introduces an
isolated pole in $W_{n+1}(z)$ which invalidates the use of Eq.(\ref{eq:Cauchy}).
Notice, however, that even in this case an orthogonal transformation of bath variables
into linear-chain modes can still be introduced to define a number of 
chains of effective modes, one for each interval where $J_{0}(\omega)>0$. 

\begin{figure}
\noindent \begin{centering}
\includegraphics[width=0.9\columnwidth]{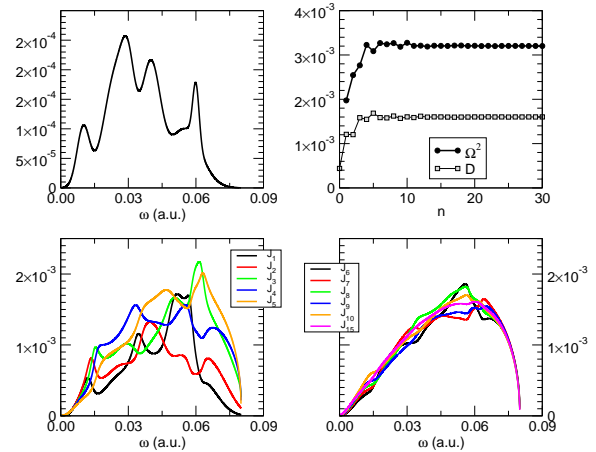}
\par\end{centering}

\caption{\label{fig:multipeaked}(Color online) Deconvolution of the highly structured SD
$J_{0}$ shown in the upper left panel, with $\omega_{R}=0.08$ a.u..
Top right: effective mode parameters up to $n=30$. Bottom: the sequence
$J_{n}$ for $n=1,\ldots,15$.}

\end{figure}

\emph{Numerical results. }The procedure described above for the determination
of the sequence $\{J_{n}\}_{n\in\mathbb{N}}$ of effective SD can
be easily implemented numerically, relying on the recurrence relation
Eq.(\ref{eq:recursive-W*}) for the SD Cauchy transforms%
\footnote{Results are numerically indistinguishable from those obtained by applying
the recursion of Eq.(\ref{eq:recursive-J*}) or by computing the eigenfrequencies
of the residual bath discretized at each step. %
}, the only necessary input being the initial SD $J_{0}$ and the cutoff
$\omega_{R}$. To show the effectiveness of the method and rapidity of
convergence we consider the numerical results for a couple of representative
SDs, with a frequency cutoff $\omega_{R}$ fixed in
such a way that $J_{0}(\omega_{R})\thickapprox0$. As a first example
we consider \begin{equation}
J_{0}(\omega)=\frac{d_{0}^{2}\gamma\omega}{(\omega^{2}-\omega_{0}^{2})^{2}+\gamma^{2}\omega^{2}},\label{eq:Garg}\end{equation}
which is the effective SD felt by a Brownian
particle coupled to a harmonic oscillator of frequency $\omega_{0}$
which in turn interacts with an Ohmic bath\citep{Garg1985a}. This
coupling scheme is evident from the results of Fig.\ref{fig:Garg}
where $J_{1}$, plotted in the middle panel, appears to be Ohmic,
as can also be checked analytically. The sequence then very rapidly
becomes indistinguishable from the Rubin SD given by Eq.\eqref{eq:Rubin}
for the chosen cutoff $\omega_{R}$, as can also be seen from the
ratio $\Omega_{n}^{2}/D_{n}\rightarrow2$. As a second example we
consider a highly structured, multipeaked SD as plotted in Fig.\ref{fig:multipeaked}.
It is clear from the figure that also in this case convergence is
quite fast, and the limiting Rubin SD is obtained after few (say 10-15)
iterations.%

\emph{Conclusions. }We have presented a recursive procedure to recast
the non-Markovian dynamics of a Brownian particle, interacting with
a bath characterized by an arbitrary SD, into the Markovian dynamics
of an enlarged set of variables including effective modes of the reservoir
coupled to a quasi-Ohmic residual SD. The approach provides an explicit
analytic relationship among successive residual SD, which can be easily
evaluated numerically starting from an arbitrary (gapless) initial
SD. These results pave the way for an efficient general treatment
of quantum dissipation in the presence of arbitrarily complex environments.

\bibliographystyle{apsrev}

\end{document}